\documentclass[11pt, draftclsnofoot, letterpaper, onecolumn, twoside]{IEEEtran}

\usepackage{epsfig,amsfonts, verbatim}
\usepackage{cite}

\newcommand{\dsp}{\displaystyle}
\newcommand{\ra}{\rightarrow}
\newcommand{\Ra}{\stackrel{w.}{\longrightarrow}}

\newtheorem{theorem}{Theorem}
\newtheorem{lemma}{Lemma}
\newtheorem{definition}{Definition}

\newtheorem{corollary}{Corollary}

\newcommand{\cG}{\mathcal{G}}
\newcommand{\cL}{\mathcal{L}}
\newcommand{\cN}{\mathcal{N}}
\newcommand{\cF}{\mathcal{F}}
\newcommand{\cS}{\mathcal{S}}
\newcommand{\bP}{\mathbb{P}}
\newcommand{\bE}{\mathbb{E}}
\newcommand{\bD}{\mathbb{D}}
\newcommand{\bC}{\mathbb{C}}
\newcommand{\cP}{\mathcal{P}}
\newcommand{\cI}{\mathcal{I}}

\newcommand{\Xb}{\bar{X}}
\newcommand{\cX}{\mathbb{X}}
\newcommand{\cXb}{\bar{\mathbb{X}}}

\newcommand{\Qt}{\tilde{Q}}

\newcommand{\Qb}{\bar{Q}}

\newcommand{\Ab}{\bar{A}}
\newcommand{\Db}{\bar{D}}
\newcommand{\Wb}{\bar{W}}
\newcommand{\Tb}{\bar{T}}
\newcommand{\Ib}{\bar{I}}
\newcommand{\Qtb}{\bar{\tilde{Q}}}

\newcommand{\xb}{\bar{x}}
\newcommand{\ab}{\bar{a}}

\title{Novel Architectures and Algorithms for Delay Reduction in
Back-pressure Scheduling and Routing}

\author{Loc~Bui, R.~Srikant, Alexander~Stolyar\thanks{L.~Bui
and R.~Srikant are with ECE Dept. and CSL, University of Illinois
at Urbana-Champaign. A.~Stolyar is with Bell Labs, Alcacel-Lucent, NJ.
Emails: locbui@ifp.uiuc.edu, rsrikant@illinois.edu, stolyar@research.bell-labs.com}}

\begin{document}

\maketitle

\begin{abstract}
The back-pressure algorithm is a well-known throughput-optimal
algorithm. However, its delay performance may be quite poor
even when the traffic load is not close to network capacity due
to the following two reasons. First, each node has to maintain
a separate queue for each commodity in the network, and only
one queue is served at a time. Second, the back-pressure
routing algorithm may route some packets along very long
routes. In this paper, we present solutions to address both of
the above issues, and hence, improve the delay performance of
the back-pressure algorithm. One of the suggested solutions
also decreases the complexity of the queueing data structures
to be maintained at each node.
\end{abstract}

\section{Introduction}
\label{sec:intro}

Resource allocation in wireless networks is complicated due to
the shared nature of wireless medium. One particular allocation
algorithm called the \emph{back-pressure algorithm} which
encompasses several layers of the protocol stack from MAC to
routing was proposed by Tassiulas and Ephremides, in their
seminal paper \cite{taseph92}. The back-pressure algorithm was
shown to be \emph{throughput-optimal}, i.e., it can support any
arrival rate vector which is supportable by any other resource
allocation algorithm. Recently, it was shown that the
back-pressure algorithm can be combined with congestion control
to fairly allocate resources among competing users in a
wireless network \cite{linshr04, sto05, sto06, erysri05,
erysri06, neemodli05}, thus providing a complete resource
allocation solution from the transport layer to the MAC layer.
While such a combined algorithm can be used to perform a large
variety of resource allocation tasks, in this paper, we will
concentrate on its application to scheduling and routing.

Even though the back-pressure algorithm delivers maximum throughput
by adapting itself to network conditions, there are several issues
that have to be addressed before it can be widely deployed in
practice. As stated in the
original paper \cite{taseph92}, the back-pressure algorithm requires
centralized information and computation, and its computational
complexity is too prohibitive for practice. Much progress has been
made recently in easing the computational complexity and deriving
decentralized heuristics. We refer the interested reader to
\cite{joolinshr08,akyetal08} and references within for some recent
results along these lines. We do not consider complexity or
decentralization issues in this paper; our proposed solutions can be
approximated well by the solutions suggested in the above papers.

Besides complexity and decentralization issues which have
received much attention recently, the back-pressure algorithm
can also have poor delay performance. To understand that, we
consider two different network scenarios: one in which the
back-pressure algorithm is used to adaptively select a route
for each packet, and the other in which a flow's route is
chosen upon arrival by some standard multi-hop wireless network
routing algorithm such as DSR or AODV and the back-pressure
algorithm is simply used to schedule packets. We refer to the
first case as \emph{adaptive-routing} and the second case as
\emph{fixed-routing}, respectively.

We first discuss networks with fixed routing. The back-pressure
algorithm assigns a weight to each flow on each link. The
weight is equal to the flow's queue backlog at the transmitter
of the link minus the flow's queue backlog at the receiver. The
weight of a link is equal to the maximum weight of any flow
that uses the link. The back-pressure algorithm then selects a
schedule which maximizes the sum of the weights of the links
included in the schedule. Under such an algorithm, for a link
to be scheduled, its weight should be slightly larger than
zero. Now, let us consider a flow that traverses $K$ links, and
use an informal argument to show why it is very intuitive that
the flow's total queue accumulation along its route should grow
quadratically with the route length. The queue length at the
destination for this flow is equal to zero. The queue length at
the first upstream node from the destination will be some
positive number, say, $\epsilon.$ The queue length at the
second upstream node from the destination will be even larger
and for the purposes of obtaining insight, let us say that it
is $2\epsilon.$ Continuing this reasoning further, the total
queue length for the flow will be
$\epsilon(1+2+\ldots+K)=\Theta(K^2).$ Thus, the total backlog
on a path is intuitively expected to grow quadratically in the
number of hops. On the other hand, suppose a fixed service rate
is allocated to each flow on each link on its path, then the
queue length at each hop will be roughly $O(1)$ depending on
the utilization at that link. With such a fixed service rate
allocation, the total end-to-end backlog should then grow
linearly in the number of hops. However, such an allocation is
possible only if the packet arrival rate generated by each flow
is known to the network a priori. One of the contributions of
this paper is to use counters called \emph{shadow queues}
introduced in \cite{buisristo08} to allocate service rates to
each flow on each link in an adaptive fashion without knowing
the set of packet arrival rates.

We will also show that the concept of shadow queues can reduce
the number of real queues maintained at each node
significantly. In particular, we will show that it is
sufficient to maintain per-neighbor queues at each node,
instead of per-flow queues required by the back-pressure
algorithm in the case of fixed routing. In large networks, the
number of flows is typically much larger compared to the number
of neighbors of each node, thus using per-neighbor queues can
result in significant reduction in implementation complexity. A
different idea to reduce the number of queues at each node has
been proposed in \cite{yinsritow08}, but the implementation
using shadow queues has the additional benefit of delay
reduction.

Next, we discuss the case of adaptive routing. The back-pressure
algorithm tends to explore many routes to find sufficient capacity in
the network to accommodate the offered traffic. Since the goal of the
algorithm is to maximize throughput, without considering Quality of
Service (QoS), back-pressure based adaptive routing can result in
very long paths leading to unnecessarily excessive delays. In this
paper, we propose a modification to the back-pressure algorithm which
forces it to first explore short paths and only use long paths when
they are really needed to accommodate the offered traffic. Thus,
under our proposed modification, the back-pressure algorithm
continues to be throughput-optimal, but it pays attention to the
delay performance of the network. We also refer the reader to a
related work in \cite{yinshared09} where the authors use the same
cost function as us, but their formulation is different and hence
their solution is also different.

%The rest of the paper is organized as follows. In
%Section~\ref{sec:sysmodel}, we present our model of the wireless
%network that will be used in subsequent sections. In
%Section~\ref{sec:shadow}, we consider networks with fixed routing
%and present the shadow queue algorithm. We also establish the
%stability of the network under per-neighbor queueing.  In
%Section~\ref{sec:performance}, we analyze the delay performance of
%the back-pressure algorithm with fixed routing and compare it to the
%performance of the shadow algorithm using simulations. In
%Section~\ref{sec:minhop}, we present an adaptive routing algorithm
%that combines the benefits of the traditional back-pressure
%algorithm with min-resource routing, and show through simulations
%that the new algorithm results in significant reduction in the queue
%backlog. Concluding remarks are provided in
%Section~\ref{sec:conclusions}.

\section{System Model}

Let us consider a network modeled by a graph, $\cG = (\cN,
\cL),$ where $\cN$ is the set of nodes and $\cL$ is the set of
links. We assume that time is slotted, with a typical time slot
denoted by $t.$ If a link $(n,m)$ is in $\cL,$ then it is
possible to transmit packets from node $n$ to node $m$ subject
to the interference constraints which will be described
shortly.

We let $\cF$ be the set of flows that share the network
resources. Packets of each flow enter the network at one node,
travel along multiple hops (which may or may not
pre-determined), and then exit the network at another node. For
each flow $f \in \cF,$ let $b(f)$ denote the begin (entering)
node, and $e(f)$ denote the end (exiting) node of flow $f.$

We define a valid schedule $\pi = \left( c_1^\pi, c_2^\pi,
\ldots, c_{|\cL|}^\pi \right)$ to be a set of link rates
(measured in terms of number of packets) that can be
simultaneously supported. Note that due to the interference
between links, for each $\pi,$ some $c_l^\pi$ could be zero.
Moreover, we make a natural and nonrestrictive assumption that
if $\pi$ is a valid schedule, then if we replace any subset of
its components by zeros, the modified schedule is valid as
well. We also assume that $c_l^\pi$ is upper-bounded by some
$c_{max}$ for any $\pi$ and $l.$ Let $\Gamma$ be the set of all
possible valid schedules, and $co(\Gamma)$ denote the convex
hull of $\Gamma.$

%We further assume that the interference is {\em local}, i.e.,
%there exists a constant $\kappa$ such that for any link $l,$
%any other link which is further than $\kappa$ hops from $l$
%will not interfere with $l.$ We call $\kappa$ the {\em
%interference-spread factor}. For example, for wired networks
%(i.e., there is no interference between links), $\kappa = 0;$
%for the primary interference model (node-exclusive interference
%model), $\kappa = 1;$ and for the two-hop interference model,
%$\kappa = 3.$ We let $c_{min}$ denote the minimum rate that any
%link is able to transmit when all of its interfering links are
%shut off.

If the routes of flows are not predetermined, i.e., when
\emph{adaptive routing} is used, then the {\em capacity region}
$\Lambda$ of the network is defined as the set of all flow
rates which are supportable by the network. Tassiulas and
Ephremides \cite{taseph92} have shown that $\lambda = \left\{
\lambda_f \right\}_{f \in \cF} \in \Lambda$ if
\begin{itemize}
\item there exists a $\mu = \left\{ \mu_{nm}\right\}_{(n,m)
    \in \cL} \in co(\Gamma),$
\item for any link $(n,m) \in \cL,$ there exists some
    allocation $\left\{ \mu^f_{nm} \right\}_{f \in \cF}$
    such that $ \mu_{nm} = \sum_{f \in \cF} \mu^f_{nm} ~,$
    and
\item for any node $n \in \cN,$ for all flows $f \in \cF,$
    $n \neq e(f),$
\[
\lambda_f \cI_{\{n = b(f)\}} + \sum_{(k,n)} \mu^f_{kn}
= \sum_{(n,m)} \mu^f_{nm}.
\]
\end{itemize}
The term $\mu^f_{nm}$ can be interpreted as the long-term
average rate that link $(n,m)$ allocates to serve flow $f.$
Note that the equation in the third bullet above is simply the
law of flow conservation.

Now, if the routes of flows are predetermined, i.e., when
\emph{fixed routing} is used, then for each $f \in \cF,$ let $L(f)$
denote the set of links forming the route of $f.$ The {\em
capacity region} $\Lambda$ of the network is defined as the set
of all flow rates which are supportable given a set of flows
and their corresponding routes. In the case of fixed routing,
$\lambda = \left\{ \lambda_f \right\}_{f \in \cF} \in \Lambda$
if there exists a $\mu = \left\{ \mu_l \right\}_{l \in \cL} \in
co(\Gamma)$ such that
$$
\sum_{f: l \in L(f)} \lambda_f \leq \mu_l, \qquad \forall l \in \cL.
$$

The traffic in the network can be \emph{elastic} or
\emph{inelastic}. If the traffic is \emph{inelastic}, i.e., the
flows' rates are fixed (and within the capacity region), then
the goal is to route/schedule the traffic through the network
while ensuring that the queues in the network are stable. If
the traffic is \emph{elastic}, then the goal is to allocate the
network's resources to all flows in some fair manner. More
precisely, suppose that each flow has a utility function
associated with it. The utility function of flow $f,$ denoted
by $U_f(\cdot),$ is defined as a function of the data rate
$x_f$ sent by flow $f,$ and assumed to be concave and
nondecreasing. The goal, in the case of elastic traffic, is to
determine the optimal solution to the following resource
allocation problem:
\begin{eqnarray}
&\max& \sum_{f \in \cF} U_f (x_f) \label{eqn:num}\\
&\mbox{s.t.}& x \in \Lambda, \nonumber
\end{eqnarray}
where $\Lambda$ is the {\em capacity region} described above.

It has been shown that, for \emph{inelastic} traffic, the
back-pressure algorithm is \emph{throughput-optimal}.
Furthermore, for \emph{elastic} traffic, a joint congestion control and
back-pressure routing/scheduling algorithm has been shown to be
able to solve the resource allocation problem (\ref{eqn:num}).
However, as we mentioned in Section~\ref{sec:intro},
the delay performance of such algorithms can be quite poor.
In the subsequent sections, we describe our architectures and
algorithms in detail.

\section{The Shadow Algorithm}\label{sec:shadow}

In this section, we consider networks with fixed routing, and
propose an architecture to reduce delays and reduce the number
of queues maintained at each node. The main idea is to use a
fictitious queueing system called the \emph{shadow queueing}
system to perform flow control and resource allocation in the
network while using only a single physical FIFO queue for each
outgoing link (also known as per-neighbor queueing) at each
node. The idea of shadow queues was introduced in \cite{buisristo08},
but the main goal there was to
extend the network utility maximization framework for wireless
networks to include multicast flows. However, one of the main
points of this work is to show that shadow queues can be useful
even in networks with unicast flows only for the purpose of
delay reduction. Further, the idea of using per-neighbor
queueing and establishing its stability is new here.

\subsection{Description}
\label{sec:shadow_algo}

The traditional back-pressure algorithm requires the queue
length of every flow that passes through a node to perform
resource allocation. The main idea of the shadow algorithm is
to decouple the storage of this information from the queueing
data structure required to store packets at each node. The
details of the shadow algorithm are described as follows.

\noindent\textbf{Queues and Counters:} At each node, instead of
keeping a separate queue for each flow as in the back-pressure
algorithm, a FIFO (first-come first-served) queue is maintained
for each outgoing link. This FIFO queue stores packets for all
flows going through the corresponding link. When a node
receives a packet, it looks at the packet's header: if the node
is not the final destination of that packet, it will send the
packet to the FIFO queue of the next-hop link; otherwise, it
will deliver the packet to the upper layer. We let $P_{nm}[t]$
denote the length of the queue maintained at link $(n,m)$ and
at the beginning of time slot $t.$

Each node maintains a separate \emph{shadow} queue (i.e., a
counter) for each flow going through it. Let $\Qt_n^f[t]$ be
the length of the shadow queue (i.e., the value of the counter)
of flow $f$ at node $n$ at the beginning of time slot $t.$ The
shadow queues and real queues are updated according to the
scheduling algorithm described next. Note that each node still
needs to keep a separate shadow queue for every flow going
through it, but these are just counters, not actual physical
queues. A counter is much easier to implement than a physical
queue.

\noindent\textbf{Back-pressure scheduling using the shadow
queue lengths:} At time slot $t,$
\begin{itemize}
\item Each link looks at the maximum \emph{shadow}
    differential backlog of all flows going through that
    link:
\begin{equation}
w_{nm}[t] = \max_{f: (n,m) \in L(f)} \left( \Qt^f_n[t] - \Qt^f_m[t]
\right). \label{eqn:shadow_diffbacklog}
\end{equation}
\item Back-pressure scheduling:
\begin{equation}
\pi^*[t] = \max_{\pi \in \Gamma} \sum_{(n,m)} c^\pi_{nm} w_{nm}[t].
\label{eqn:shadow_wirelesssched}
\end{equation}
\item A schedule $\pi^*=( c_1^\pi, c_2^\pi, \ldots,
    c_{|\cL|}^\pi )$ is interpreted by the network as
    follows: link $(n,m)$ transmits $c_{nm}^\pi$ shadow
    packets from the shadow queue of the flow whose
    differential backlog achieves the maximum in
    (\ref{eqn:shadow_diffbacklog}) (if the shadow queue has
    fewer than $c_{nm}^\pi$ packets, then it is emptied);
    link $(n,m)$ also transmits as many real packets as
    shadow packets from its real FIFO queue. Again, if the
    number of real packets in the queue is less than the
    number of transmitted shadow packets, then all the real
    packets are transmitted.
\end{itemize}
We recall that shadow queues are just counters. The action of
``transmitting shadow packets'' is simply the action of
updating the counters' values. In other words, ``transmitting''
$k$ shadow packets from $\Qt^f_n$ to $\Qt^f_m$ means that we
subtract $k$ from $\Qt^f_n$ and add $k$ to $\Qt^f_m.$ From the
above description, it should be clear that the shadow packets
can be interpreted as permits which allow a link to transmit.
Unlike the traditional back-pressure algorithm, the permits are
associated with just a link rather than with a link and a flow.

\noindent\textbf{Congestion control at the source:} At time
slot $t,$ the source of flow $f$ computes the rate at which it
injects packets into the ingress {\em shadow} queue as follows:
\begin{equation}\label{eqn:congestion control}
x_f[t] = \min \left\{ U_f^{'-1} \left( \frac{\Qt^f_{b(f)}[t]}{M} \right),
x_{max} \right\},
\end{equation}
where $x_{max}$ is an upper-bound of the arrival rates, and $M$
is a positive parameter. The source also generates real traffic
at rate $\beta x_f[t]$ where $\beta$ is a positive number less
than $1.$ If $x_f$ and $\beta x_f$ are not integers, the actual
number of shadow and real packets generated can be random
variables with these expected values. Since the shadow packets
are permits that allow real-packet transmission, from basic
queueing theory, it follows that the actual packet arrival rate
must be slightly smaller than the shadow packet arrival rate to
ensure the stability of real queues. The parameter $\beta$ is
chosen to be less than $1$ for this purpose. As we will see
later in simulations, the queue backlog in the network would be
smaller for smaller values of $\beta.$

The above description of the shadow algorithm applies to
elastic traffic. For inelastic traffic, the same shadow
algorithm can be used without congestion control. To ensure
stability of the real queues, if the real arrival rate of an
inelastic flow is $\lambda_f,$ the shadow arrival rate for this
flow must be larger than $\lambda_f.$ For example, if we wish
to make the shadow arrival rate larger than the real arrival
rate by a factor of $(1+\epsilon),$ it can be accomplished as
follows: for every real packet arrival, generate a shadow
packet. Generate an additional shadow packet for each real
packet with probability $\epsilon.$ This procedure ensures that
the shadow arrival rate will be $(1+\epsilon)$ times the real
arrival rate. For the algorithm to be stable, the set of
arrival rates $\{\lambda_f(1+\epsilon)\}_f$ must lie in the
interior of capacity region.

Alternatively, the shadow algorithm for inelastic traffic can
be implemented slightly differently if we are willing to
tolerate packet loss: fix the shadow arrival rate for each flow
and regulate the arrival rate of real packets to be a fraction
$\beta$ of the shadow arrival rate. For example, if the rate of
shadow arrivals in a time slot is $\lambda_f,$ then one can
inject real packets according to a Poisson distribution of mean
$\beta \lambda_f.$ The real packets could be stored in a queue
at its ingress node, and drained at this rate to inject into
the network. If the mean real arrival rate is larger than
$\beta$ times the mean shadow arrival rate, then the real
packet buffer at the edge of the network will overflow leading
to packet loss. Packet loss is unavoidable for inelastic flows
unless the arrival rate is less than the capacity that the
network is willing to allocate to the flow. The shadow arrival
rate in this case should be thought of as the
network-designated capacity for a flow.

We note that the concept of shadow queues here is different
from the notion of virtual queues used in \cite{gibkel99} for
the Internet and in \cite{erysri05} for wireless networks. In
networks with virtual queueing systems, the arrival rates to
both the real and virtual queues are the same, but the virtual
queue is drained at a slower rate than the real queue. Instead,
here the arrival rates to the real queues are slightly smaller
than the arrival rates to the corresponding shadow queues. This
subtle difference is important in that it allows us to use
per-neighbor FIFO queues and prove stability in a multihop
wireless network in the next section.

\subsection{Stability of the shadow algorithm}
\label{sec:shadow_perf}

In this subsection, we establish the optimality and stability
of the real and shadow queues. First, we note that the
optimality of the resource allocation and the stability of
shadow queues follow from previous results in the literature.
In particular, we have the following theorem.

\begin{theorem} \label{thm:shadowstability}
The shadow-queue-based congestion control and scheduling
algorithms described in Section~\ref{sec:shadow_algo} above
asymptotically achieve the optimal rate allocation, i.e.,
\begin{equation}
\label{eq-optimality}
\lim_{T\ra\infty} \frac{1}{T} \sum_{t=0}^{T-1} \bE[x[t]] =
x^*+O(1/M),
\end{equation}
where $x^*$ is the optimal solution to (\ref{eqn:num}).
Furthermore, the {\em shadow} queues are stable in the sense
that the Markov chain of shadow queues $\Qt[t]$ is positive
recurrent and the steady-state expected values of the shadow
queue lengths are bounded as follows:
$$\sum_{n,f} \bE(\Qt_n^f[\infty]) =O(M).$$
\end{theorem}

\begin{IEEEproof}
The proof of this theorem was
presented in \cite{erysri06, sto05, neemodli05}, and hence,
is omitted here. $\hfill$
\end{IEEEproof}

The remaining goal is to prove the stability of the real
queues. Note that the sources are sending real traffic with
smaller rates than shadow traffic, and we know that the shadow
queues are stable. However, it does not automatically mean that
the real queues are stable as well, since each of them is an
aggregated FIFO queue storing packets for all flows going
through its corresponding link. Fortunately, we can apply
results from the stochastic networks literature to establish
the following result.

\begin{theorem}\label{thm:fifostability}
The process describing the joint evolution of both shadow and
real queues,
$$\left( \left(\Qt^f_n[t]\right)_{f \in \cF, n \in
\cN}; \left(P_{nm}[t]\right)_{(n,m) \in \cL} \right),$$
is an irreducible, aperiodic, positive recurrent Markov chain.
Therefore, the real FIFO queues are also stable.
\end{theorem}

The proof is based on the fluid limit approach and a result by
Bramson \cite{bra96}. In his paper, Bramson proved that fluid models
of Kelly-type FIFO queueing networks are stable as long as the
nominal load on each server is strictly less than its capacity. Thus,
the basic idea of the proof is as follows. The random process
describing the behavior of {\em shadow} queues, under the joint
congestion control and scheduling algorithm (running on the shadow
system), is positive recurrent (as specified in
Theorem~\ref{thm:shadowstability}). Therefore, the {\em average}
service rate on each network {\em link} that the shadow algorithm
yields is strictly greater than the nominal load of the link due to
the thinning of actual traffic; moreover, the (random) cumulative
amount of service provided on each link up to time $t$ satisfies the
functional strong law of large numbers, as $t$ goes to infinity. As a
result, if we take the {\em fluid limit} of the process describing
real FIFO queues, it has {\em exactly same form as if each network
link would have constant, non-time-varying capacity (equal to the
average rate provided by the shadow algorithm)}. Then, this fluid
limit is stable by the results of \cite{bra96}, which implies
stability of the process of real queues. The proof's details are
presented in Appendix~\ref{app:proof_sketch} just for the purpose of
completeness.

Note that the real traffic throughput will always be slightly
smaller than the optimal solution to (\ref{eqn:num}), but this
difference from the optimal solution can be made arbitrarily
small by adjusting the parameter $\beta.$

\section{Performance Comparison: Back-Pressure Algorithm versus
the Shadow Algorithm}\label{sec:performance}

In this section, we compare and contrast the performances of
the traditional back-pressure algorithm and the shadow
algorithm for networks with fixed routing.

\subsection{Elastic traffic under the traditional
back-pressure algorithm}\label{sec:line elastic}

\begin{figure}
  \centering
  \centerline{\psfig{file=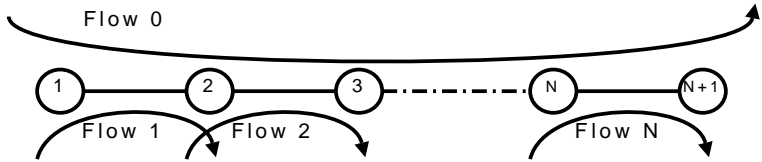,width=4in}}
  \caption{The linear network with $N$ links.}
\label{fig:linear_net}
\end{figure}

We present simple calculations in this section to get some feel
for the performance of the traditional back-pressure algorithm
when it is used with congestion control. Deriving expressions
for the queue backlog for general-topology networks seems quite
hard, so we confine our discussions to the case of a linear
network with $N$ links as in Figure~\ref{fig:linear_net}.
There are $N+1$ flows sharing this network: one flow (indexed
$0$) goes through all $N$ links, and $N$ other flows (indexed
$1$ to $N$) where each of them goes through each link. The
capacity of each link is $c,$ and for simplicity, we assume
that there is no interference between links.

Let $x_i$ and $U_i(\cdot)$ denote the rate and the utility
function flow $i,$ respectively. The network utility
maximization problem we would like to solve is as follows:
\begin{eqnarray*}
&\max& \sum_{i=0}^N U_i(x_i) \\
&\mbox{s.t.}& x_0 ~\leq~ \mu_{0,1} , \\
&& \mu_{0,i} ~\leq~ \mu_{0,i+1}, ~ i = 1, \ldots, N-1, \\
&& x_i + \mu_{0,i} ~\leq~ c, ~ i = 1, \ldots, N,
\end{eqnarray*}
where $\mu_{0,i}$ is the resource that link $i$ allocates to
serve flow $0.$

If the utility is logarithmic (proportional fairness), i.e.,
$U_i(x) = \log(x),$ then one can easily compute the optimal
rates and optimal queue lengths (which are the Lagrange
multipliers) for the above optimization problem as follows:
\begin{eqnarray}
&& x^*_0 = \mu^*_{0,1} = \ldots = \mu^*_{0,N} = \frac{c}{N+1}, \nonumber \\
&& x^*_1 = \ldots = x^*_N = \frac{N c}{N+1}, \nonumber \\
&& q^*_i = q^*_{0,i} - q^*_{0,i+1} = \frac{N+1}{Nc}, ~ i = 1,
\ldots, N, \qquad \label{eqn:ex_diff1}
\end{eqnarray}
where $q^*_i$ and $q^*_{0,i}$ are the optimal queue lengths
maintained at node $i$ for flow $i$ and flow $0$, respectively.
Then, the end-to-end total queue length for flow $0$ is
\[
\sum_{i=1}^N q^*_{0,i} = \frac{N+1}{Nc} \sum_{i=1}^N i =
\frac{(N+1)^2}{2c} = \Theta\left(N^2\right).
\]

For a more general class of utility functions which model a
large class of fairness concepts \cite{mowal00},
\[
U_i(x) = \frac{x^{1-\alpha}}{1-\alpha}, \quad \alpha > 0,
\]
we still have similar results:
\begin{eqnarray}
&& x^*_0 = \mu^*_{0,1} = \ldots = \mu^*_{0,N} =
\Theta\left(N^{-1/\alpha}\right),
\nonumber\\
&& x^*_1 = \ldots = x^*_N = \Theta(1), \nonumber\\
&& q^*_i = q^*_{0,i} - q^*_{0,i+1} = \Theta(1), ~ i = 1, \ldots, N,
\qquad \label{eqn:ex_diff2}
\end{eqnarray}
which again lead to $\sum_{i=1}^N q^*_{0,i} =
\Theta\left(N^2\right).$ As mentioned in the Introduction
section, if a fixed rate (larger than its arrival rate) is
allocated to each flow, then the total queue length in this
network is expected to increase as the order of $N$ instead of
$N^2.$

\subsection{Inelastic traffic under the traditional
back-pressure algorithm} \label{sec:sim2}

In the previous subsection, we showed that the combined
back-pressure and congestion control algorithm for elastic
traffic can lead to quadratic end-to-end queueing delay in
terms of the number of hops. It is interesting to see whether
such a behavior can also be observed in the case of inelastic
traffic, i.e., the flows' rates are fixed, and the traditional
back-pressure algorithm is used. The following theorem
establishes an upper-bound on the end-to-end queue backlog for
any flow.

\begin{theorem} \label{thm:delay_uppperbound}
Consider a general topology network accessed by a set of flows
with fixed routes. Let $K_{max}$ be the maximum number of hops
in the route of any flow, i.e., $K_{max} = \max_f |L(f)|.$
Suppose the arrival rate vector $\lambda$ is such that, for
some $\epsilon>0$, $(1+\epsilon)\lambda$ lies in the interior
of the capacity region of the network. Assume that the arrival
processes of the flows are independent of each other,
independent from time slot to time slot, and have finite second
moments. Then, the expected value of the sum of queue lengths
(in steady-state) along the route of any flow $f$ is bounded as
follows:
%grows at most quadratically in $K_{max}$, i.e.,
\[
\bE \left[ \sum_{n \in R(f)} Q^f_n [\infty]\right] ~\leq~
\frac{1+\epsilon}{\epsilon} \frac{b}{\lambda_f} |\cF| K_{max}^2
%O\left(K_{max}^2\right)
~,~ \forall f \in \cF,
\]
where constant $b>0$ depends only on $c_{max}$.

\end{theorem}

\begin{IEEEproof}
The proof is presented in Appendix~\ref{app:delay_proof}.
\end{IEEEproof}

While the above result is only an upper bound, it suggests the
quadratic growth of the total flow queue length on the flow
route length. The simulation results shown next validate such
quadratic growth.

\subsection{Simulation results for inelastic traffic}

To illustrate the queue length behavior under back-pressure
algorithm in the case of inelastic traffic, we simulate the
linear network in Figure~\ref{fig:linear_net}. We choose $N =
40,$ i.e., the network has $41$ nodes and $40$ links, with no
interference between links. Each link has capacity $10,$ i.e.,
it can transmit up to $10$ packets per time slot. Let
$\lambda_0$ be the fixed rate of flow $0,$ and $\lambda_1$ be
the fixed rate of flows $1, 2, \ldots, 40.$ We know that the
back-pressure algorithm will stabilize the network as long as
$\lambda_0 + \lambda_1 < 10.$ We let the sources send shadow
traffic at fixed rates $\lambda_i,$ and send real traffic at a
slightly smaller rate $\beta \lambda_i,$ with $\beta \in (0,
1)$.

\begin{figure}
  \centering
  \centerline{\psfig{file=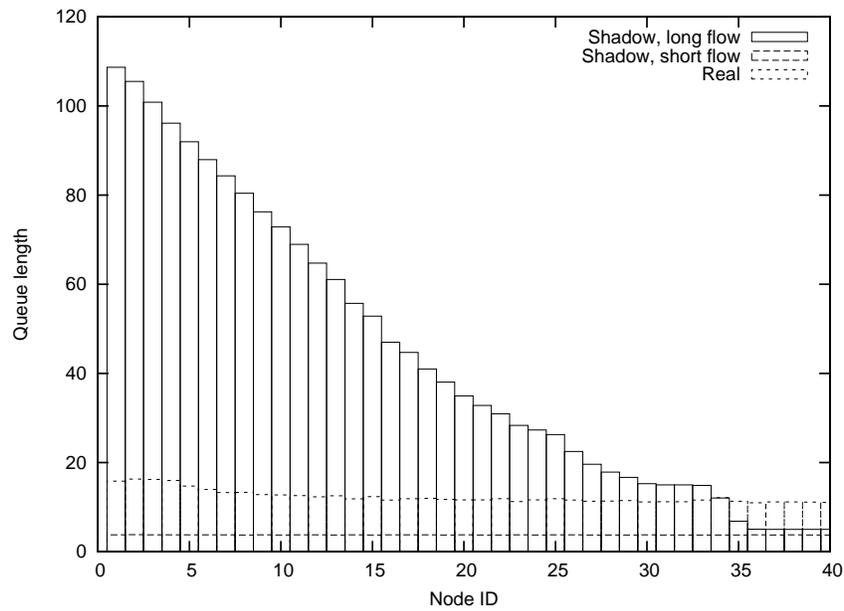,width=4.5in}}
  \caption{The queue lengths at each node in the linear network
  in Figure~\ref{fig:linear_net}. The solid-line boxes are the
  lengths of shadow queues of flow $0$ (the long flow) maintained
  at each node. The dash-line boxes are the shadow queue lengths of flows
  $i$, $i = 1, \ldots, 40,$ (the short flows) at node $i,$ respectively.
  Finally, the dot-line boxes are the real FIFO queue lengths at each node.}
\label{fig:simres2}
\end{figure}

Figure~\ref{fig:simres2} shows the mean queue lengths of all
queues maintained at each node when $\lambda_0 = 5$ and
$\lambda_1 = 2.5.$ The value of $\beta$ here is $0.99.$ We see
that the shadow queue lengths of flow $0$ increase nearly
linearly when going from the end node to the begin node, which
leads to a quadratic growth (in terms of the number of hops) of
the end-to-end queue backlog. Moreover, we also see that the
real FIFO queue lengths are significantly reduced, even with a
small amount thinning of traffic ($1\%$).

\subsection{Simulation results for elastic traffic}
\label{sec:sim1}

In this subsection, we investigate the performance of the
shadow algorithm with elastic traffic in a network with a more
complicated topology than a line.
\begin{figure}
  \centering
  \centerline{\psfig{file=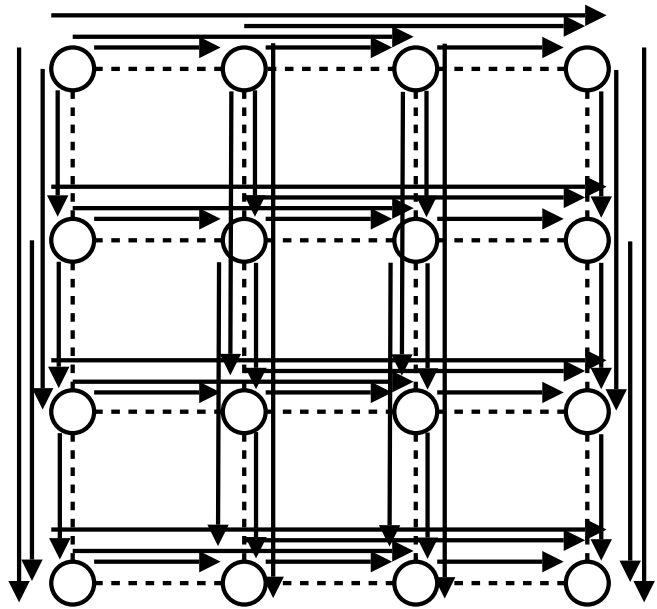,width=3.5in}}
  \caption{A grid network with $16$ nodes, $24$ links, and $48$ flows.
  Links and flows are represented by dash lines and solid arrows,
  respectively.}
\label{fig:square_net}
\end{figure}
In particular, we consider a grid network as shown in
Figure~\ref{fig:square_net}. We assume that all flows have a
logarithmic utility function, i.e., $U_f(x_f)=\log x_f$ for all
$f.$ The network has $16$ nodes (represented by circles) and
$24$ links (represented by dash lines). We assume a simple
one-hop interference model under which a matching in the graph
represents a valid schedule. Each link has a capacity of $10,$
i.e., it can transmit up to $10$ packets in one time slot if
scheduled. There are $48$ flows (represented by arrows) sharing
this network.

\begin{figure}
  \centering
  \centerline{\psfig{file=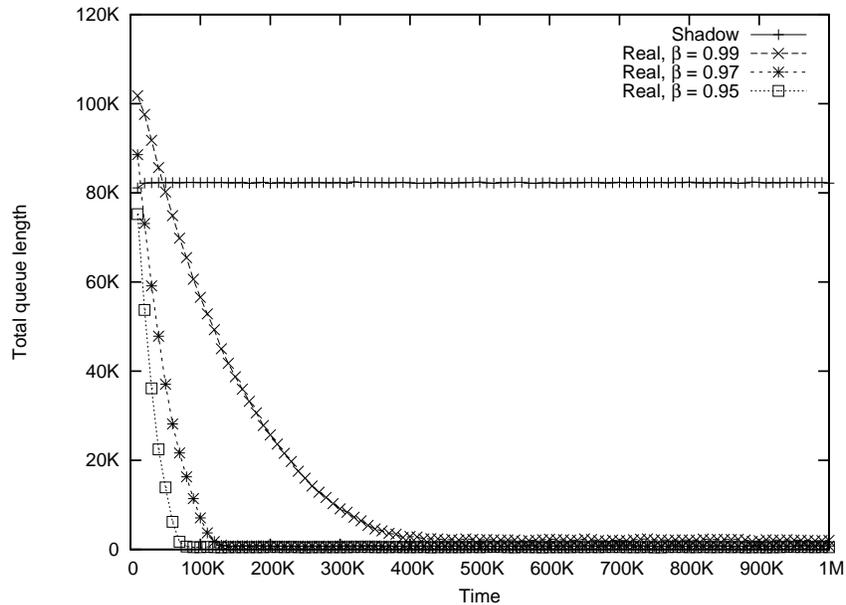,width=4.5in}}
  \caption{The evolutions of total shadow queue length and total real
  queue lengths with different values of $\beta$ over time.}
\label{fig:simres1}
\end{figure}

We implement the shadow algorithm as described in
Section~\ref{sec:shadow_algo} with the parameter $M=1000.$ In
Figure~\ref{fig:simres1}, we plot the evolution of total shadow
queue length and total real queue length for several values of
parameter $\beta$ (the total queue length is the sum of all
queue lengths in the network). Note that the shadow queue
length is also the queue length of traditional back-pressure
scheduling without the shadow algorithm. The figure indicates
that the total real queue length with the shadow algorithm
decreases dramatically compared to the traditional
back-pressure algorithm. Since the shadow queue length is very
large compared to the real queue lengths, it is hard to see the
actual values of the real queue lengths in the figure, so we
present some numbers from the simulations here: after a half
million time slots, the total shadow queue length is around
$82\,000$ while the total real queue lengths are only about
$2000,$ $800,$ and $500,$ when $\beta$ is $0.99,$ $0.97,$ and
$0.95,$ respectively. Thus, significant gains in performance
can be realized at the expense of a small loss in throughput
(represented by the parameter $1-\beta$). Note that the
traditional back-pressure algorithm can perform poorly due to
many reasons: (i) As in Section~\ref{sec:line elastic}, if the
number of hops for a flow is large, then the queue backlog can
increase quadratically. (ii) The choice of the parameter $M$ in
the congestion control algorithm (see Equation
(\ref{eqn:congestion control})) can lead to queue backlogs of
the order of $M$ (see the upper bound in
Theorem~\ref{thm:shadowstability} and simulation results in
\cite{erysri05}). (iii) A separate queue is maintained for each
destination. The shadow algorithm solves all of these problems
at once by ``reserving" capacity between each
source-destination pair, i.e., for each flow.

\section{Min-Resource Routing using Back-Pressure Algorithm}
\label{sec:minhop}

In this section, we consider wireless networks where each
flow's route is not pre-determined, but is adaptively chosen by
the back-pressure algorithm for each packet. As mentioned in 
the Introduction section, the back-pressure algorithm
explores all paths in the network and as a result may choose
paths which are unnecessarily long and may even contain loops,
thus leading to poor performance. We address this problem by
introducing a cost function which measures the total amount of
resources used by all the flows in the network. Specifically,
we add up traffic load on all links in the network and use this
as our cost function. In the case of inelastic flows, the goal
then is to minimize this cost subject to network capacity
constraints. Due to the nature of the cost function, in a
network with links of equal capacity, shorter hop paths will be
preferred over longer hop paths.

In the case of elastic flows, one can maximize the sum of flow
utilities minus a weighted function of the cost described
above, where the weight provides a tradeoff between network
utility and resource usage. Since the solutions to both
problems are similar, we only present the inelastic case here.

\subsection{Description}
\label{sec:minhop_des}

Given a set of packet arrival rates that lie within the
capacity region, our goal is to find the routes for flows such
that as few network resources as possible are used. Thus, we
formulate the following optimization problem:
\begin{eqnarray}
&\min& \sum_{(n,m)} \mu_{nm} \label{eqn:minhop_opt}\\
&s.t.& \lambda_f \cI_{\{n = b(f)\}} + \sum_{(k,n)} \mu^f_{kn}
~\leq~ \sum_{(n,m)} \mu^f_{nm} , \quad
\forall f \in \cF, n \in \cN, \nonumber\\
&& \{\mu_{nm}\}_{(n,m)\in \cal L} ~\in~ co(\Gamma), \nonumber
\end{eqnarray}
where $\mu^f_{nm}$ is the rate that link $(n,m)$ allocates to
serve flow $f,$ i.e., $\mu_{nm} = \sum_f \mu^f_{nm},$  and
$\lambda_f$ is the fixed rate of flow $f.$
%The parameter $\gamma > 0$ does not
%affect the solution to the above problem but as we will see later,
%it influences the transient behavior of the routing algorithm
%designed to solve the problem. We can use Lagrange multipliers to
%rewrite the optimization problem as
%\[
%\min_{\mu \in co(\Gamma)} \sum_f \lambda_f \cI_{\{n = b(f)\}}
%q^f_{b(f)} -\sum_f \sum_{(n,m)} \left( q^f_n - q^f_m - \gamma
%\right) \mu^f_{nm} ,
%\]
%where $q^f_n$ are the Lagrange multipliers. Since the first
%term in the objective function does not show up in the
%constraint, the above optimization is equivalent to
%\[
%\max_{\mu \in co(\Gamma)} \sum_f \sum_{(n,m)} \left( q^f_n - q^f_m -
%\gamma \right) \mu^f_{nm},
%\]
%which has the form of back-pressure algorithm. In fact, one can
%follow standard steps in the proof of
%Theorem~\ref{thm:shadowstability} to show that the following
%modified back-pressure algorithm will lead to the optimal solution
%of the min-resource routing problem (\ref{eqn:minhop_opt})
%asymptotically:
An algorithm that asymptotically solves the min-resource
routing problem (\ref{eqn:minhop_opt}) is as follows. (It is a
special case of the algorithm in \cite{sto05}, where the
scaling parameter $1/M$ is called $\beta$.)

\noindent\textbf{Min-resource routing by back-pressure:} At
time slot $t,$
\begin{itemize}
\item Each node $n$ maintains a separate queue of packets
    for each destination $d$; its length is denoted
    $Q^d_n[t]$. Each link is assigned a weight
% equal to the maximum differential backlog
%    minus a constant $\gamma :$
\begin{equation}
w_{nm}[t] = \max_{d \in D}
\left( \frac{1}{M}Q^d_n[t] - \frac{1}{M}Q^d_m[t] - 1
\right), \label{eqn:minhop_diff1}
\end{equation}
where $M>0$ is a parameter (having the same meaning as
earlier in this paper).
\item Scheduling/routing rule:
\begin{equation}
\pi^*[t] = \max_{\pi \in \Gamma} \sum_{(n,m)} \pi_{nm} w_{nm}[t].
\label{eqn:minhop_sched}
\end{equation}
\item If the schedule $\pi^*$ says, for example, to send
    $c_{nm}^\pi$ packets over link $(n,m),$ then link
    $(n,m)$ transmits up to $c_{nm}^\pi$ packets from the
    queue $Q^d_n$ to $Q^d_m$ for the destination $d$
    achieving the maximum in (\ref{eqn:minhop_diff1}).
\end{itemize}

Note that the above algorithm does not change if we replace the
weights in (\ref{eqn:minhop_diff1}) by the following, re-scaled ones:
\begin{equation}
w_{nm}[t] = \max_{d \in D}
\left( Q^d_n[t] - Q^d_m[t] - M
\right). \label{eqn:minhop_diff}
\end{equation}
Therefore, compared with the traditional back-pressure
scheduling/routing, the only difference is that each link
weight is equal to the maximum differential backlog {\em minus
parameter $M$}. ($M=0$ reverts the algorithm to traditional.)

The performance of the stationary process which is ``produced''
by the algorithm with fixed parameter $M$ is within $O(1/M)$ of
the optimal (analogously to (\ref{eq-optimality})):
$$
\left| ~\bE \left[\sum_{(n,m)} \pi_{nm}^*[t]\right] - \sum_{(n,m)} \mu_{nm}^*
~\right|= O(1/M),
$$
where $\mu^*$ is an optimal solution to (\ref{eqn:minhop_opt}).
However, larger $M$ means larger $O(M)$ queues and slower
convergence to the (nearly optimal) stationary regime. On the
other hand, ``too small'' $M$ results in a stationary regime
being ``too far'' from optimal, and queues being large for that
reason. Therefore, a good value for $M$ for a practical use
should be neither too large nor too small. Our simulations
confirm these intuitions.

\subsection{Simulation results}

\begin{figure}
  \centering
  \centerline{\psfig{file=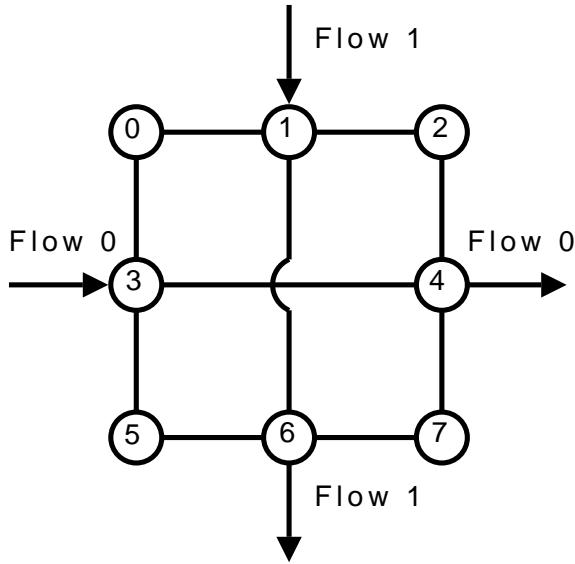,width=3in}}
  \caption{A network with $8$ nodes, $10$ links, and $2$ flows.
  Each link has a capacity of $10.$ Flow $0$ enters at node $3$
  and exits at node $4,$ while flow $1$ enters at node $1$ and exits
  at node $6.$}
\label{fig:dia_net}
\end{figure}

We ran a simulation for a network with $8$ nodes, $10$ links, and $2$
flows as in Figure~\ref{fig:dia_net}.
We assume the node-exclusive
spectrum sharing interference model, i.e., each node can only
communicate with at most one other node in any time slot. Each link
has a capacity of $10$ when activated. Flow $0$ enters the network at
node $3$ and exits at node $4,$ while flow $1$ enters at node $1$ and
exits at node $6.$ Note that the flows' routes have not been
prespecified, and the described above algorithm with parameter $M$ is
used.

We fix each flow's rate at value $\lambda.$ It is easy to see that
under the node-exclusive spectrum sharing interference model, the
back-pressure algorithm can stabilize the network as long as $\lambda
< 10.$ The arrival processes are Poisson, i.e., the number of
arrivals for each flow at each time slot is a Poisson random variable
with mean $\lambda.$ Each simulation run was $1$ million time-slots
long and $40$ such runs were performed. The results reported are
averaged over these $40$ runs.

Table~\ref{tab:simres3}
\begin{table}
\caption{The link's rate allocation for network in
Figure~\ref{fig:dia_net} when each flow's rate is $\lambda = 5.0.$}
\label{tab:simres3} \centering
\begin{tabular}{|@{}c@{}|c|c|c|c|c|c|}
\multicolumn{7}{c}{} \\
\hline
& \multicolumn{2}{|c|}{$M=0$} & \multicolumn{2}{|c|}{$M=10$}
& \multicolumn{2}{|c|}{$M=20$} \\
\hline
Link & Rate for & Rate for & Rate for & Rate for & Rate for & Rate for \\
$(n,m)$ & flow $0$ & flow $1$ & flow $0$ & flow $1$ & flow $0$ & flow $1$ \\
\hline
$(0,1)$ & 1.9492 & 1.9671 & 0.0000 & 0.0000 & 0.0000 & 0.0000 \\
$(1,2)$ & 1.9759 & 1.5622 & 0.0000 & 0.0000 & 0.0000 & 0.0000 \\
$(0,3)$ & 1.9055 & 2.0058 & 0.0000 & 0.0000 & 0.0000 & 0.0000 \\
$(1,6)$ & 0.0563 & 2.2417 & 0.0000 & 4.9998 & 0.0000 & 5.0001 \\
$(2,4)$ & 1.4913 & 2.4595 & 0.0000 & 0.0000 & 0.0000 & 0.0000 \\
$(3,4)$ & 2.2504 & 0.0466 & 4.9993 & 0.0000 & 4.9996 & 0.0000 \\
$(3,5)$ & 1.5551 & 1.9881 & 0.0000 & 0.0000 & 0.0000 & 0.0000 \\
$(4,7)$ & 1.2590 & 2.0853 & 0.0000 & 0.0000 & 0.0000 & 0.0000 \\
$(5,6)$ & 2.3592 & 1.5055 & 0.0000 & 0.0000 & 0.0000 & 0.0000 \\
$(6,7)$ & 2.0094 & 1.2535 & 0.0000 & 0.0000 & 0.0000 & 0.0000 \\
\hline
\end{tabular}
\end{table}
shows the rate allocation of each link to each flow when the value of
$\lambda$ is fixed at $5.0$ and for $M = 0,$ $10,$ and $20.$ Note
that $M = 0$ corresponds to the traditional back-pressure algorithm.
We see that the traditional back-pressure algorithm uses all links in
the network, while our modified back-pressure algorithm (with $M =
10$ or $M = 20$) essentially uses only link $(3,4)$ for flow $0$ and
link $(1,6)$ for flow $1$ (which are the min-resource routes for
these flows).

%\begin{table}[t]
%\centering
%\begin{tabular}{|@{}c@{}|c|c|c|c|c|c|}
%\hline
%& \multicolumn{2}{|c|}{$\gamma=0$} & \multicolumn{2}{|c|}{$\gamma=10$}
%& \multicolumn{2}{|c|}{$\gamma=20$} \\
%\hline
%Link & Rate for & Rate for & Rate for & Rate for & Rate for & Rate for \\
%$(n,m)$ & flow $0$ & flow $1$ & flow $0$ & flow $1$ & flow $0$ & flow $1$ \\
%\hline
%$(0,1)$ & 0.4602 & 0.4559 & 0.0000 & 0.0000 & 0.0000 & 0.0000 \\
%$(1,2)$ & 0.4752 & 0.3685 & 0.0000 & 0.0000 & 0.0000 & 0.0000 \\
%$(0,3)$ & 0.4612 & 0.4493 & 0.0000 & 0.0000 & 0.0000 & 0.0000 \\
%$(1,6)$ & 0.0177 & 8.2218 & 0.0000 & 9.0007 & 0.0000 & 8.9993 \\
%$(2,4)$ & 0.4686 & 0.3648 & 0.0000 & 0.0000 & 0.0000 & 0.0000 \\
%$(3,4)$ & 8.1302 & 0.0201 & 8.9998 & 0.0000 & 8.9995 & 0.0000 \\
%$(3,5)$ & 0.4117 & 0.4661 & 0.0000 & 0.0000 & 0.0000 & 0.0000 \\
%$(4,7)$ & 0.4023 & 0.3205 & 0.0000 & 0.0000 & 0.0000 & 0.0000 \\
%$(5,6)$ & 0.4164 & 0.4613 & 0.0000 & 0.0000 & 0.0000 & 0.0000 \\
%$(6,7)$ & 0.4043 & 0.3169 & 0.0000 & 0.0000 & 0.0000 & 0.0000 \\
%\hline
%\end{tabular}
%\caption{The link's rate allocation for network in
%Fig.~\ref{fig:dia_net} when each flow's rate is $\lambda =
%9.0.$} \label{tab:simres3}
%\end{table}

\begin{figure}
  \centering
  \centerline{\psfig{file=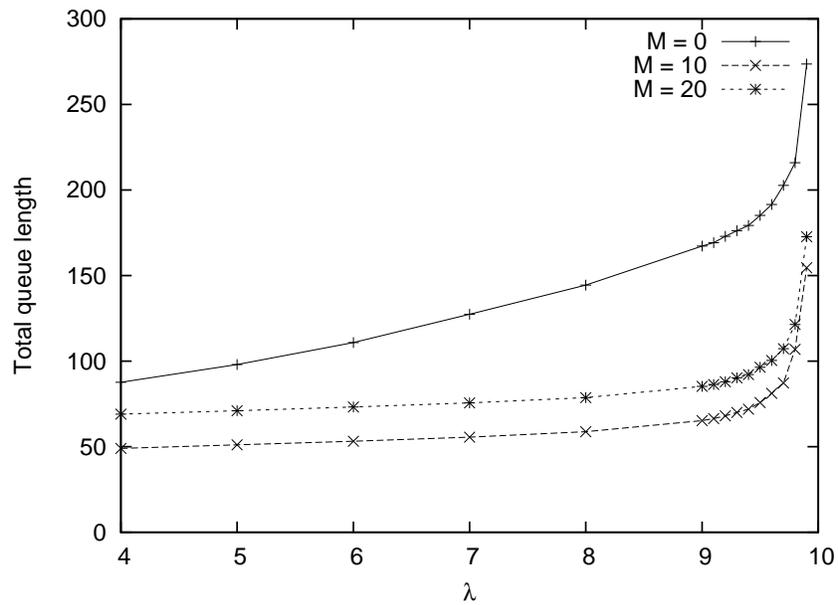,width=4.5in}}
  \caption{The sum of all queue lengths of the network in
Figure~\ref{fig:dia_net} corresponding to various values of
$\lambda$.} \label{fig:simres4}
\end{figure}

We then turn our attention to the queue backlog (the sum of all queue
lengths) in the network. Figure~\ref{fig:simres4}
shows the queue backlog in the network corresponding to various
values of $\lambda.$ We see that the queue backlog of our modified
back-pressure algorithm with $M=10$ is significantly smaller than
that of the traditional back-pressure algorithm. However, when $M$ is
increased to $20,$ the delay performance gets slightly worse. This
result confirms our observation about the trade-off in choosing the
value of $M$ which is discussed at the end of
Section~\ref{sec:minhop_des}.

%\begin{table}[t]
%\centering
%\begin{tabular}{|c|c|c|c|}
%\hline
%Values of $\lambda$ & $\gamma=0$ & $\gamma=10$ & $\gamma=20$ \\
%\hline
%9.0 & 169.038 & 65.435 & 85.291 \\
%9.1 & 169.937 & 66.471 & 86.364 \\
%9.2 & 174.121 & 68.048 & 87.938 \\
%9.3 & 176.219 & 69.974 & 90.255 \\
%9.4 & 177.880 & 72.579 & 92.177 \\
%9.5 & 189.150 & 76.363 & 96.366 \\
%9.6 & 191.510 & 82.504 & 100.414 \\
%9.7 & 200.712 & 88.548 & 110.234 \\
%9.8 & 215.937 & 110.593 & 121.396 \\
%9.9 & 253.409 & 154.570 & 162.730 \\
%10.0 & 4440.350 & 3928.780 & 4113.240 \\
%\hline
%\end{tabular}
%\caption{The sum of all queue lengths of the network in
%Fig.~\ref{fig:dia_net} corresponding to various values of
%$\lambda$.} \label{tab:simres4}
%\end{table}

\section{Conclusions}\label{sec:conclusions}

In this paper, we have proposed a new shadow architecture to improve
the delay performance of back-pressure scheduling algorithm. The
shadow queueing system allows each node to maintain a single FIFO
queues for each of its outgoing links, instead of keeping a separate
queue for each flow in the network. This architecture not only
reduces the queue backlog (or, equivalently, delay by Little's law)
but also reduces the number of actual physical queues that each node
has to maintain. Next, we proposed an algorithm that forces the
back-pressure algorithm to use the minimum amount of network
resources while still maintaining throughput optimality. This
results in better delay performance compared to the traditional
back-pressure algorithm.

We presented the shadow algorithm for the case of fixed
routing, i.e., the route for each flow is fixed. The shadow
algorithm can also be used in the case of adaptive routing, but
a node cannot use just one FIFO queue for each neighbor. If one
still maintains a separate queue for each destination at each
node, then the extension of the shadow algorithm to the case of
adaptive routing is straightforward. On the other hand, it
would be interesting to study if a single per-neighbor FIFO
queue can be maintained even in the case of adaptive routing.
This is an interesting topic for future research.

\appendices

\section{Proof of Theorem~\ref{thm:fifostability}}
\label{app:proof_sketch}

In this appendix, we provide details of the proof of
Theorem~\ref{thm:fifostability}. First, recall the result from
Theorem~\ref{thm:shadowstability} that
\begin{equation}
\lim_{T\ra\infty} \frac{1}{T}
\sum_{t=0}^{T-1} \bE[x[t]] = x^*(\epsilon),
\label{eqn:timeavgxb}
\end{equation}
where $x^*(\epsilon)$ is within $\epsilon$-boundary of the
optimal solution $x^*$ and $\epsilon$ can be made arbitrarily
small by increasing $M.$ To simplify the notations, from now
on, we will drop $\epsilon$ in $x^*(\epsilon).$ In other words,
we will use the notation $x^*$ for the $\epsilon$-approximate
optimal solution.

From the above result, the following convergence results can be
established.

\begin{lemma} \label{lem:arrivalproc}
For every flow $f \in \cF,$
$$
\frac{1}{T} \sum_{t=0}^{T-1} a_f[t]
~\stackrel{T \ra \infty}{\longrightarrow}~
\beta x^*_f \qquad m.s.,
$$
i.e., the time average of real packet arrivals also converges
to the optimal solution.
\end{lemma}

\begin{IEEEproof}

Consider any flow $f.$ We have the sequence of flow rates
$\left\{ x_f[0], x_f[1], \ldots \right\}$ and the sequence of
generated shadow packets $\left\{ a_f[0], a_f[1], \ldots
\right\}.$ Note that given the sequence of flow rates $\left\{
x_f[t] \right\}_{t=0}^{\infty},$ $a_f[t]$'s are independent
Poisson random variables with means $\beta x_f[t]$'s. For
simplicity, we drop the subscript $f$ in the notations within
the scope of this proof.

Let $\xb[t] := \bE[x[t]],$ and $\ab[t] := \bE[a[t]] = \bE[\beta
x[t]] = \beta \xb[t].$
%From Theorem~\ref{thm:shadowstability}, we know that:
%\begin{equation}
%\lim_{T\ra\infty} \frac{1}{T} \sum_{t=0}^{T-1} \xb[t] = x^*.
%\label{eqn:timeavgxb}
%\end{equation}
%Also, since the Markov chain of shadow queues are positive recurrent,
%by the ergodicity theorem,
%\begin{equation}
%\lim_{T\ra\infty} \frac{1}{T} \sum_{k=0}^{T-1} x[k] = \xb[\infty]
%\qquad a.s. \label{eqn:timeavgx}
%\end{equation}
We have that
\begin{eqnarray*}
\bE \left[ \left( \frac{1}{T}\sum_{i=1}^T a[i] - \beta x^* \right)^2\right]
&=& \frac{1}{T^2} \bE \left[ \left( \sum_{i=1}^T (a[i] - \beta  x^*)
\right)^2\right] \\
&=& \frac{1}{T^2} \sum_{i=1}^T \bE \left[ (a[i] - \beta x^*)^2 \right] \\
&& +~ \frac{1}{T^2} \sum_{i=1}^T \sum_{j=1, j\neq i}^T
\bE \left[ (a[i] - \beta x^*) (a[j] - \beta x^*) \right].
\end{eqnarray*}
Now, for $i \neq j,$
\begin{eqnarray*}
Cov(a[i], a[j]) &=& \bE \left[(a[i] - \ab[i]) (a[j] - \ab[j]) \right] \\
&=& \bE \left[ \bE \left[\left.(a[i] - \ab[i]) (a[j] - \ab[j]) \right|
\{ x[k] \}_{k=0}^\infty \right] \right] \\
&=& 0,
\end{eqnarray*}
i.e., $a[i]$ and $a[j]$ are uncorrelated. The last equality is
due to the fact that $a[i]$ and $a[j]$ are independent given
the sequence $\{x_k\}_{k=0}^\infty.$ Hence,
\begin{eqnarray*}
\bE \left[ (a[i] - \beta x^*) (a[j] - \beta x^*) \right]
&=& \bE \left[ (a[i] - \ab[i]) (a[j] - \ab[j]) \right]
+ (\ab[i] - \beta x^*) (\ab[j] - \beta x^*) \\
&=& (\ab[i] - \beta x^*) (\ab[j] - \beta x^*).
\end{eqnarray*}
Furthermore, $ \bE \left[ (a[i] - \beta x^*)^2 \right] = \bE
\left[ (a[i] - \ab_i)^2 \right] + (\ab[i] - \beta x^*)^2.
$
Therefore,
\begin{eqnarray*}
\bE \left[ \left( \frac{1}{T}\sum_{i=1}^T a[i] - \beta x^* \right)^2\right]
&=& \frac{1}{T^2} \sum_{i=1}^T Var(a[i]) + \frac{1}{T^2}
\sum_{i=1}^T \sum_{j=1}^T (\ab[i] - \beta x^*) (\ab[j] - \beta x^*) \\
&=& \frac{1}{T^2} \sum_{i=1}^T Var(a[i]) +
\left[ \frac{1}{T} \sum_{i=1}^T (\ab[i] - \beta x^*)\right]^2.
\end{eqnarray*}
Since $Var(a[i])$ is finite, the first term in the right-hand
side will vanish as $T$ goes to infinity. The second term also
vanishes as $T$ goes to infinity due to (\ref{eqn:timeavgxb}).
Therefore,
\begin{eqnarray*}
\lim_{T \ra \infty} \bE \left[ \left( \frac{1}{T}\sum_{i=1}^T a[i]
- \beta x^* \right)^2\right] &=&  0.
\end{eqnarray*}
\end{IEEEproof}

\begin{lemma} \label{lem:serviceproc}
Recall that $\pi^*[t]$ is the outcome of the scheduling
algorithm (\ref{eqn:shadow_wirelesssched}) at every time step
$t.$ Then, for every link $l \in \cL,$
$$
\frac{1}{T} \sum_{t=0}^{T-1} \pi^*_l[t] ~\stackrel{T \ra \infty}
{\longrightarrow}~ \mu^*_l \qquad a.s.
$$
for some $\mu^*$ such that $ \dsp \sum_{f: l \in L(f)}
\lambda_f \leq \mu^*_l, ~ \forall l \in \cL.$ \\ \\
In other words, the outcome of the scheduling algorithm
converges to a set of link rates that can support the given set
of flow arrival rates.
\end{lemma}

\begin{IEEEproof}
Since the Markov chain of shadow
queues is positive recurrent, the proof follows the ergodic
theorem and the fact that $\pi^*$ is upper-bounded.
\end{IEEEproof}

To be consistent with \cite{bra96}, we introduce the concept of
{\em packet class}. Each flow $f$ consists of $|L(f)|$ packet
classes; each class going through one link in the route of $f.$
We let $\cS$ denote the set of all packet classes. In other
words, there is a bijection mapping a pair $(f, l),$ $f \in
\cF,$ $l \in L(f),$ to a packet class $s \in \cS.$ Clearly,
$|\cS| = \sum_{f \in \cF} |L(f)|.$

For each flow $f \in \cF,$ let $\Phi(f)$ be the set of packet
classes belonging to $f.$ For each link $l \in \cL,$ let $C(l)$
be the set of packet classes going through $l.$ Conversely, for
each packet class $s \in \cS,$ let $f(s)$ be the corresponding
flow (i.e., $s \in \Phi(f(s))$), and $l(s)$ be the
corresponding link.

Let $H$ denote the {\em constituency matrix} with size $|\cL|
\times |\cS|$:
\begin{eqnarray*}
H_{l,s} &=& \left\{
\begin{array}{ll}
1 & \mbox{if } s \in C(l) ,\\
0 & \mbox{otherwise.}
\end{array}
\right.
\end{eqnarray*}

Also, let $R$ be the {\em routing matrix} with size $|\cS|
\times |\cS|$:
\begin{eqnarray*}
R_{s,u} &=& \left\{
\begin{array}{ll}
1 & \mbox{if $f(s) = f(u)$ and $u$ is the next
    hop of $s$ in the route of } f, \\
0 & \mbox{otherwise.}
\end{array}
\right.
\end{eqnarray*}

Next, let $E_s(t)$ denote the total {\em external} arrivals of
packet class $s$ up to time $t.$ Thus,
\begin{eqnarray*}
E_s(t) &=& \left\{
\begin{array}{ll}
\dsp \sum_{k=0}^{t-1} a_f[k] & \mbox{if $s$ is the first hop of } f(s),\\
0 & \mbox{otherwise.}
\end{array}
\right.
\end{eqnarray*}

Also, we define the arrival rates corresponding to packet
classes:
\begin{eqnarray*}
\lambda_s &=& \left\{
\begin{array}{ll}
\dsp x^*_f & \mbox{if $s$ is the first hop of } f(s),\\
0 & \mbox{otherwise.}
\end{array}
\right.
\end{eqnarray*}

We then extend the definition of $E_s(t)$ for continuous time
as follows: for each time $t \in \Re^+,$ $E_s(t) := E_s \left(
\lfloor t \rfloor\right).$ Hence, $E_s(t)$ is right continuous
having left limits in time.

Recall that $\pi^*[t]$ is the outcome of the scheduling
algorithm at time slot $t.$ Now, for each $t \in \Re^+,$ we let
$M_l(t):= M_l\left( \lfloor t \rfloor\right) = \sum_{k=0}^{t-1}
\pi^*_l[k]$ denote the total amount of service (in terms of
number of packets that {\em can be} transmitted) of link $l$ up
to time $t.$ Now, for each $s \in \cS,$ let us define $m_s(t)
:= M_{l(s)}(t),$ and define
\begin{equation}
M(t) := diag \left( \frac{\lfloor t \rfloor}{m_1(t)}, \frac{\lfloor t \rfloor}
{m_2(t)}, \ldots, \frac{\lfloor t \rfloor}{m_{|\cS|}(t)} \right).
\label{eqn:servicetime}
\end{equation}

Similarly, let us define $A_s(t) = A_s \left( \lfloor t
\rfloor\right)$ as the total arrivals, and $D_s(t) = D_s \left(
\lfloor t \rfloor\right)$ as the total departures, of packet
class $s$ up to time $t.$ Thus,
\begin{equation}
A_s(t) = E_s(t) + \sum_{u \in \cS} D_u(t) R_{u,s}.
\label{eqn:arrival1}
\end{equation}

Let $Q_s(t)= Q_s \left( \lfloor t \rfloor\right)$ be the number
of packets of packet class $s$ which are waiting to be served.
Then,
\begin{equation}
Q_s(t) = Q_s(0) + A_s(t) - D_s(t).
\end{equation}

Recall that $P_l(t)= P_l \left( \lfloor t \rfloor\right)$ is
the length of FIFO queue at link $l$ at time $t.$ Thus,
\begin{equation}
P_l(t) = \sum_{s \in C(l)} Q_s(t) = \sum_s H_{l,s} Q_s(t).
\end{equation}

Now, we define
\begin{itemize}
\item $T_s(t)$ as the amount of time that the server at
    link $l(s)$ has spent serving packet class $s$ in
    $[0,t];$
\item $I_l(t)$ as the amount idle time of the server at
    link $l$ during $[0,t];$
\item $W_l(t)$ as the immediate workload at the server of
    link $l,$ measured in units of time.
\end{itemize}

Then we have the following equations:
\begin{equation}
\sum_s H_{l,s} T_s(t) + I_l(t) = t
\end{equation}
\begin{equation}
W_l(t) = \frac{\lfloor t \rfloor}{M_l(t)} \sum_s H_{l,s} \left( A_s(t) + Q_s(0)\right)
- \sum_s H_{l,s} T_s(t), \label{eqn:workload1}
\end{equation}
and the fact that $I_l(t)$ can only increase when $W_l(t) = 0,$
i.e., if $I_l(t_2) > I_l(t_1)$ then $W_l(t) = 0$ for some $t
\in [t_1, t_2].$

We can rewrite the above equations
(\ref{eqn:arrival1})-(\ref{eqn:workload1}) in vector form to
get the following set of equations which describes the
evolution of the system:
\begin{eqnarray}
&& A(t) = E(t) + R^T D(t) \label{eqn:arrival} \\
&& Q(t) = Q(0) + A(t) - D(t)  \label{eqn:queue} \\
&& P(t) = H Q(t)  \label{eqn:FIFOqueue} \\
&& H T(t) + I(t) = e t  \label{eqn:service} \\
&& W(t) = H M(t) [ A(t) + Q(0) ] - H T(t)  \label{eqn:workload} \\
&& I_l(t) \mbox{ can only increase when } W_l(t) = 0, ~ l \in \cL ,
\label{eqn:idletime}
\end{eqnarray}
where $M(t)$ is defined in (\ref{eqn:servicetime}) and $e = [1,
1, \ldots, 1]^T.$ Additionally, we have that
\begin{eqnarray}
&& M(t) D(t) \leq T(t) \leq M(t) (D(t)+ e )  \label{eqn:HOLcond} \\
&& D_s\left( t + W_{l(s)}(t) \right) = Q_s(0) + A_s(t)
\label{eqn:FIFOcond}
\end{eqnarray}
where Equation (\ref{eqn:HOLcond}) comes from the fact that
each class has at most one packet being served at any time, and
Equation (\ref{eqn:FIFOcond}) comes from the FIFO property of
the real queues.

Note that $E_s(t),$ $M_l(t),$ $A_s(t),$ $D_s(t),$ $Q_s(t),$ and
$W_l(t)$ are right continuous having left limits, while
$T_s(t)$ and $I_l(t)$ are continuous in time. We also assume
that $A(0) = D(0) = T(0) = I(0) = 0.$

Let us define
\begin{eqnarray*}
X(t) &:=& ( A(t), D(t), Q(t), W(t), T(t), I(t), \Qt(t) ),
\end{eqnarray*}
where $\Qt_s(t)= \Qt_s \left( \lfloor t \rfloor\right)$ is the
shadow queue of class $s.$ Then $X(t)$ is a Markov process.
Furthermore, $\Qt(t)$ and $(Q(t),\Qt(t))$ are themselves Markov
processes. By Theorem~\ref{thm:shadowstability}, we know that
$\Qt(t)$ is positive recurrent.

We now describe the fluid model of the system. The set of fluid
model equations is as follows:
\begin{eqnarray}
&& A(t) = \beta \lambda t + R^T D(t) \label{eqn:f_arrival} \\
&& Q(t) = Q(0) + A(t) - D(t)  \label{eqn:f_queue} \\
&& P(t) = H Q(t)  \label{eqn:f_FIFOqueue} \\
&& H T(t) + I(t) = e t  \label{eqn:f_service} \\
&& W(t) = H M [ A(t) + Q(0) ] - H T(t)  \label{eqn:f_workload} \\
&& I_l(t) \mbox{ can only increase when } W_l(t) = 0, ~ l \in \cL
\label{eqn:f_idletime} \\
&& T(t) = M D(t)  \label{eqn:f_HOLcond} \\
&& D_s\left( t + W_{l(s)}(t) \right) = Q_s(0) + A_s(t) ,
\label{eqn:f_FIFOcond}
\end{eqnarray}
where $ M^* = diag \left( \frac{1}{m^*_1}, \frac{1}{m^*_2},
\ldots , \frac{1}{m^*_{|\cS|}}\right),$ and $m^*_s =
\mu^*_{l(s)}.$ Recall that $\mu^*$ is defined in
Lemma~\ref{lem:serviceproc} as the set of supporting link
rates. Equation (\ref{eqn:f_idletime}) means that for each $t >
0,$ whenever $W_l(t) > 0,$ there exists $\delta > 0$ such that
$I_l(t+\delta) = I_l(t-\delta),$ i.e., $I_l(\cdot)$ is constant
in $(t-\delta,t+\delta).$

\subsection{Preliminaries}

\begin{theorem}[see \cite{bil87}] \label{thm:bil87}
If random variables $Z_n$ and $Z$ satisfy that $Z_n \Ra Z,$
where the notation $(\Ra)$ denotes the convergence in
distribution (weak convergence), and if the $Z_n$ are {\em
uniformly integrable}, i.e.,
$$
\lim_{\alpha \ra \infty} \sup_n \int_{|Z_n| > \alpha} |Z_n| d\bP = 0,
$$
then $Z$ is integrable and
$$
\lim_n \bE[Z_n] = \bE[Z].
$$
\end{theorem}

Consider the sequence of scaled processes
$$
X^r(t) = \frac{1}{r} X(rt),~ t \geq 0,~ r = 1, 2, \ldots,
$$
then we have the following theorem:
\begin{theorem}[\cite{rybsto92,dai95}] \label{thm:rybsto92}
Suppose that, for any sequence of scaled processes $X^r(t)$
satisfying $\| (Q^r(0),\Qt^r(0)) \|= 1,$ $r \ra \infty,$ there
exist a subsequence $r_k \ra \infty$ and a constant $T > 0$
such that
$$
\lim_{r_k \ra \infty} \bE \|(Q^{r_k}(t),\Qt^{r_k}(t))\| = 0,
\qquad \forall ~ t \geq T.
$$
Then the queueing system is stable, in the sense that the
Markov process $(Q(t),\Qt(t))$ is positive recurrent.
\end{theorem}

\begin{corollary}
Suppose that there exists a deterministic function
$$\Xb(t)=(\Ab(t),\Db(t),\Qb(t),\Wb(t),\Tb(t),\Ib(t),
\Qtb(t))$$ such that the following conditions hold:
\begin{enumerate}[(i)]
\item For any sequence $r \ra \infty,$ there exists a
    subsequence $r_k \ra \infty$ such that $X^{r_k}(\cdot)
    \Ra \Xb(\cdot)$ as $r_k \ra \infty.$
\item For any $\Xb(t)$  satisfying $\|(\Qb(0),\Qtb(0))\| =
    1,$ there exists a $T > 0$ such that
    $\|(\Qb(t),\Qtb(t))\|=0,~ \forall t \geq T,$
\item $(Q^r(t),\Qt^r(t))$ is uniformly integrable for all
    $t > 0,$
\end{enumerate}
then the original process $(Q(t),\Qt(t))$ is positive
recurrent.
\end{corollary}

\begin{IEEEproof}

From conditions (i) and (ii), we have that
$\|(\Qb(t),\Qtb(t))\| \Ra 0$ for all $t \geq T$ as $r \ra
\infty .$ Along with condition (iii), Theorem~\ref{thm:bil87}
yields that $\lim_{n\ra \infty} \bE [\|(\Qb(t),\Qtb(t))\|] = 0,
~ \forall t \geq T.$ We then apply Theorem~\ref{thm:rybsto92}
to get the result.
\end{IEEEproof}

\subsection{Proof's details}

For an integer $d \geq 1,$ let $\bD^d[0,\infty)$ be the set of
functions $f : [0,\infty) \ra \Re^d$ that are right continuous
on $[0,\infty)$ having the left limits on $(0,\infty).$ For $t
> 0,$ we use $f(t-)$ to denote $\lim_{s \uparrow t} f(s).$ By
convention, $f(0-) = f(0).$

Let us endow the function space $\bD^d[0,\infty)$ with the
Skorohod $J_1-$topology. We now define the convergence of a
sequence of functions in $\bD^d[0,\infty)$ under that topology.
Let $\Lambda$ denote the set of strictly increasing, continuous
functions $f : \Re^+ \ra \Re^+$ such that $f(0) = 0$ and
$\lim_{t \ra \infty} f(t) = \infty.$

\begin{definition}
A sequence $\{f^n\} \subset \bD^d[0,\infty)$ is said to
converge to $f \in \bD^d[0,\infty)$ in the Skorohod topology if
for each $t > 0,$ there exists $\{\lambda^n\} \subset \Lambda$
such that
\begin{eqnarray*}
&& \lim_{n\ra \infty} \sup_{0 \leq s \leq t} | \lambda^n(s) - s | = 0 \\
&& \lim_{n\ra \infty} \sup_{0 \leq s \leq t} | f^n(\lambda^n(s)) - f(s) | = 0.
\end{eqnarray*}
\end{definition}

Next, let us define the convergence under the uniform topology.

\begin{definition}
A sequence $\{f^n\} \subset \bD^d[0,\infty)$ is said to
converge to $f \in \bD^d[0,\infty)$ uniformly on compact
intervals (u.o.c.) as $n \ra \infty$, denoted by $f^n \ra f$
u.o.c., if for each $t > 0,$
\[
\lim_{n\ra \infty} \sup_{0 \leq s \leq t} | f^n(x) - f(s) | = 0.
\]
\end{definition}

Note that $\bD^d[0,\infty)$ under the Skorohod topology is
separable, while $\bD^d[0,\infty)$ under the u.o.c. topology is
not. However, if the limit point $f$ is continuous, the two
notions of convergence are equivalent. We let $\bC^d[0,\infty)$
denote the set of continuous functions $f: [0,\infty) \ra
\Re^d.$

Now, consider the scaled process
\begin{equation}
X^r(t) = \frac{1}{r} X(rt),~ t \geq 0,~ r = 1, 2, \ldots .
\label{eqn:scaledproc}
\end{equation}
The processes $X(t)$ and $X^r(t)$ take values in
$\bD^K[0,\infty),$ where $K = 8 |\cS| + 2 |\cL|.$

\begin{lemma} \label{lem:convergence}
For any sequence $r \ra \infty,$ there exists a subsequence
$r_k \ra \infty$ such that
\begin{equation} \label{eqn:weakconvcond}
X^{r_k}(t) \Ra \Xb(t)
\end{equation}
for some $\Xb(t) \in \bD^K[0,\infty).$ Moreover, $\Xb(t)$ is
continuous with probability one.
\end{lemma}

\begin{IEEEproof}
By Lemma~\ref{lem:arrivalproc}, we have that
$$\lim_{T \ra \infty} E(T) = \beta \lambda T \qquad m.s.$$
Thus, for each $t > 0,$ $E^r(t)$ converges to $\beta \lambda t$
in probability as $r \ra \infty;$ i.e., given any $\epsilon >
0,$
$$
\lim_{r \ra \infty} \bP \left( \| E^r(t) - \beta \lambda t\| >
\frac{\epsilon}{2} \right) = 0.
$$
Furthermore, if $\| E^r(t) - \beta \lambda t\| \leq
\frac{\epsilon}{2},$ then
$$
\| E^r(t_1) - E^r(t_2) \| \leq \| E^r(t_1) - \beta \lambda t_1 \| +
\| E^r(t_2) - \beta \lambda t_2 \| + \beta \lambda |t_1 - t_2|
\leq \beta \lambda |t_1 - t_2| + \epsilon.
$$
Therefore, $E^r(t)$ is ``asymptotically Lipchitz''; i.e., for
any $\epsilon,$
$$
\bP \left( \| E^r(t_1) - E^r(t_2) \| \leq \beta \lambda |t_1 - t_2|
+ \epsilon \right) ~\ra~ 1 \qquad \mbox{as} \qquad r \ra \infty.
$$
Also, from Lemma~\ref{lem:serviceproc}, we have that for each
$l \in \cL:$
$$\lim_{T\ra\infty}M_l(T) = \mu^*_l T \qquad a.s.$$
Furthermore, the processes $D(t), Q(t), I(t), T(t), \Qt(t)$
have bounded increments. Thus, it is easily to see that
$X^r(t)$ is ``asymptotically Lipchitz''; i.e., for any
$\epsilon,$ there exists $L
> 0$ such that
$$
\bP \left( \| X^r(t_2) - X^r(t_1) \| \leq L (t_2 - t_1) + \epsilon \right)
\ra 1 \qquad \mbox{as} \qquad r \ra \infty.
$$
This implies the sequence $\left\{ X^r(t) \right\}$ is
relatively compact (ref. Corollary 3.7.4, \cite{ethkur86}),
i.e., the family of their associated probability distributions,
denoted by $\left\{\cP^r(\cdot)\right\},$ is relatively
compact. Thus, there exists a subsequence of
$\left\{\cP^r(\cdot)\right\}$ which converges to some
$\cP(\cdot)$ under the Prohorov metric. This then implies that
there exists a sub-sequence of $\left\{ X^r(t) \right\}$ which
weakly converges to some $\Xb(t)$ (ref. Theorem 3.3.1,
\cite{ethkur86}). Moreover, it follows from this weak
convergence and the asymptotic Lipchitz property that the limit
$\Xb(t)$ is continuous with probability one (ref. Theorem
3.10.2, \cite{ethkur86}).
\end{IEEEproof}

We call any $\Xb(t)$ satisfying (\ref{eqn:weakconvcond}) a {\em
fluid limit} of $X(t).$ Given a fluid limit $\Xb(t)$ and the
converging subsequence $\left\{ X^{r_k}(t) \right\},$ the
Skorohod representation theorem implies that there exist some
processes $\cXb(t)$ and $\{\cX^{r_k}(t)\}$ in a common
probability space such that $\cXb(t)$ and $\cX^{r_k}(t)$ have
the same probability distributions as $X(t)$ and $X^{r_k}(t),$
and that $\cX^{r_k}(t)$ converges to $\cXb(t)$ almost surely
under the Skorohod topology. Furthermore, since the limit point
is continuous with probability one, that convergence is in the
almost sure sense under the uniform topology, i.e.,
\begin{equation} \label{eqn:asconvcond}
\cX^{r_k}(t) \ra \cXb(t)
\qquad \mbox{u.o.c. with probability } 1.
\end{equation}

Now, let us abuse the notations by denoting
$$
\cX(t) = ( A(t), D(t), Q(t), W(t), T(t), I(t), \Qt(t)),
$$
where the components of $\cX(t)$ satisfy the set of equations
(\ref{eqn:arrival})-(\ref{eqn:FIFOcond}). Also, let
$$
\cXb(t) = (\Ab(t),\Db(t),\Qb(t),\Wb(t),\Tb(t),\Ib(t),\Qtb(t))
$$
be a corresponding fluid limit; i.e., the condition
(\ref{eqn:asconvcond}) holds as $r_k \ra \infty.$ We will show
that the components of $\cXb(t)$ satisfy the set of fluid model
equations (\ref{eqn:f_arrival})-(\ref{eqn:f_FIFOcond}), as
stated in the following lemma.

\begin{lemma}
Any fluid limit $\cXb(t)$ in (\ref{eqn:asconvcond}) satisfies
the set of fluid model equations
(\ref{eqn:f_arrival})-(\ref{eqn:f_FIFOcond}). Also, the
component $\Qtb(t)$ of $\cXb(t)$ is zero for all $t.$
\end{lemma}

\begin{IEEEproof}
First, combining
Lemmas~\ref{lem:arrivalproc}-\ref{lem:convergence}, and
(\ref{eqn:asconvcond}), we know that
\begin{eqnarray*}
\lim_{r_k \ra \infty} E^{r_k}(t) = \beta \lambda t \qquad
\mbox{u.o.c. with probability } 1, \\
\lim_{r_k \ra \infty} M^{r_k}_l(t) = \mu_l t \qquad
\mbox{u.o.c. with probability } 1.
\end{eqnarray*}
Then, it is easy to see that $\cXb(t)$ satisfies
(\ref{eqn:f_arrival})-(\ref{eqn:f_workload}),
(\ref{eqn:f_HOLcond}), and (\ref{eqn:f_FIFOcond}) since
$\cX(t)$ satisfies (\ref{eqn:arrival})-(\ref{eqn:workload}),
(\ref{eqn:HOLcond}), and (\ref{eqn:FIFOcond}), respectively. To
prove (\ref{eqn:f_idletime}) for $\cXb(t),$ we need to show
that for each $t
> 0,$ whenever $\Wb_l(t) > 0,$ then there exists $\delta > 0$ such that
$\Ib_l(t + \delta) = \Ib_l(t - \delta),$ i.e., $\Ib_l(s)$ is
flat in $(t-\delta,t+\delta).$ Suppose that $\Wb_l(t) > 0$ for
any $t > 0.$ Since $\Wb_l(t)$ is continuous, there exists a
$\delta > 0$ such that $\epsilon =
\min_{s\in(t-\delta,t+\delta)} \Wb_l(s) > 0.$ Since
$\cXb(\cdot)$ is a fluid limit, there exists a sample path
$\omega$ such that
$$
\left( W^{r_k}(\cdot,\omega), I^{r_k}(\cdot, \omega)\right)
\ra \left( \Wb(\cdot), \Ib(\cdot)\right) \qquad \mbox{u.o.c.}
$$
as $r_k \ra \infty.$ In particular, there exists an integer $N$
such that
$$
\inf_{s\in(t-\delta,t+\delta)} W^{r_k}_l(s,\omega) \geq \epsilon / 2
$$
for $r_k \geq N.$ It means that $W_l(s,\omega) > 0$ for
$s\in(r_k(t-\delta),r_k(t+\delta))$ and $r_k \geq N.$ Thus, by
(\ref{eqn:idletime}), $I_l(s, \omega)$ is flat for
$s\in(r_k(t-\delta),r_k(t+\delta))$ when $r_k \geq N,$ or
equivalently, $I^{r_k}_l(s,\omega)$ is flat for
$s\in(t-\delta,t+\delta).$ Letting $r_k \ra \infty,$ we have
that $\Ib_l(s)$ is flat for $s\in(t-\delta,t+\delta),$ and
hence we prove (\ref{eqn:f_idletime}). Finally, from the
positive recurrence of $\Qt(t)$, it is easy to see that
$\Qtb(t)$ is zero for all $t.$
\end{IEEEproof}

So the final step is to show that any solution to the set of
fluid model equations
(\ref{eqn:f_arrival})-(\ref{eqn:f_FIFOcond}) is stable. In
fact, this is true by Bramson's result \cite{bra96}. It thus
completes the proof of Theorem~\ref{thm:fifostability}.

\section{Proof of Theorem~\ref{thm:delay_uppperbound}}
\label{app:delay_proof}

Recall that $L(f)$ is the set of links forming the route of
flow $f.$ Now, we let $R(f)$ denote the set of nodes forming
the route of $f$ (and hence, $|R(f)| = |L(f)|+1$). For each
pair $(f,n)$ such that $n \in R(f),$ we abuse the notation by
letting
\begin{itemize}
\item $n+1$ denote the next node of $n$ in the route of $f$
    $(n \neq e(f));$
\item $n-1$ denote the previous node of $n$ in the route of
    $f$ $(n \neq b(f)).$
%\item $l_o = (n,n+1) \in L$ denote the outgoing link of
%    flow $f$ at node $n$ $(n \neq e(f));$
%\item $l_i = (n-1,n) \in L$ denote the incoming link of
%    flow $f$ at node $n$ $(n \neq b(f)).$
\end{itemize}

For each $n \in R(f),$ let us define
\[
\pi_{out(f,n)}[t] ~:=~ \pi^f_{(n,n+1)}[t]~, \quad n \neq e(f) ,
\]
\[
\pi_{in(f,n)}[t] := \left\{
\begin{array}{ll}
a_f[t], & n = b(f), \\
\min\left\{\pi^f_{(n-1,n)}[t], Q^f_{n-1}[t]\right\}, &
n \neq b(f).
\end{array}
\right.
\]
The queue dynamics are given by
\begin{equation}
Q^f_n[t+1] = \left( Q^f_n[t] - \pi_{out(f,n)}[t]
\right)^+ + \pi_{in(f,n)}[t]. \label{eqn:queuerule}
\end{equation}
Now, consider the Lyapunov function
\begin{eqnarray*}
V(Q) &=& \frac{1}{2} \sum_{f \in \cF} \sum_{n \in R(f)}
(Q^f_n)^2.
\end{eqnarray*}
We can rewrite the queues' dynamics (\ref{eqn:queuerule}) as
follows:
\[
Q^f_n[t+1] = Q^f_n[t] -\pi_{out(f,n)}[t] + \pi_{in(f,n)}[t]
+ u^f_n[t],
\]
where
\[
u^f_n[t] = \left\{
\begin{array}{ll}
0 &\mbox{if } Q^f_n[t] \geq \pi_{out(f,n)}[t],\\
\dsp - Q^f_n[t] + \pi_{out(f,n)}[t]
&\mbox{if } Q^f_n[t] < \pi_{out(f,n)}[t].
\end{array}
\right. \label{eqn:u}
\]
The drift of the Lyapunov function is given by
\begin{eqnarray*}
\Delta V[t] &:=&
\bE \left[ \left. V(Q[t+1]) - V(Q[t]) \right| Q[t] \right] \\
&=& \frac{1}{2} \sum_{f \in \cF} \sum_{n \in R(f)} \bE \left[ 2 Q^f_n[t]
\left( \pi_{in(f,n)}[t] - \pi_{out(f,n)}[t]\right) \right.
+ \left( \pi_{in(f,n)}[t] - \pi_{out(f,n)}[t] \right)^2\\
&& +~ 2 u^f_n[t] \pi_{in(f,n)}[t] +
\left. \left. \left( u^f_n[t]\right)^2 + 2 u^f_n[t]
\left( Q^f_n[t] - \pi_{out(f,n)}[t] \right) \right| Q[t]
\right].
\end{eqnarray*}
%It is easy to see from (\ref{eqn:u}) that
%$(\ref{eqn:driftbound2}) \leq 0.$ Further, note that
%$\lambda^f_n \geq 0,$ and $0 \leq u^f_n[t] \leq
%\pi_{out(f,n)}[t]~\forall t.$ Hence,
%\begin{eqnarray*}
%(\ref{eqn:driftbound1}) &\leq&
%\left( \pi_{in(f,n)}[t] - \pi_{out(f,n)}[t] \right)^2
%+ 2 \pi_{out(f,n)}[t] \pi_{in(f,n)}[t] \\
%&=& \left( \pi_{in(f,n)}[t]\right)^2 + \left( \pi_{out(f,n)}[t]
%\right)^2.
%\end{eqnarray*}
Recall that $ \pi_{in(f,n)}[t] = \pi_{out(f,n-1)}[t] -
u^f_{n-1}[t], ~ n \neq b(f). $ Thus, we get
\begin{eqnarray*}
\Delta V[t]&=& B_1[t] + \sum_{f \in \cF} Q^f_{b(f)}[t] \lambda_f
- \sum_{f \in \cF} \sum_{(n,m) \in L(f)} \left( Q^f_n[t] -
Q^f_m[t] \right) \bE \left[ \left. \pi^f_{nm}[t] \right| Q[t]\right] \\
&=& B_1[t] + \sum_{f \in \cF} Q^f_{b(f)}[t] \lambda_f
- \sum_{(n,m) \in \cL} \pi^{*}_{nm}[t] \max_{f:(n,m) \in L(f)}
\left( Q^f_n[t] - Q^f_m[t] \right)^+ ,
\end{eqnarray*}
where the last equality is due to the back-pressure scheduling
algorithm, and
\begin{eqnarray*}
B_1[t] &=& \frac{1}{2} \sum_{f \in \cF} \sum_{n \in R(f)} \bE \left[
\left( \pi_{in(f,n)}[t] - \pi_{out(f,n)}[t] \right)^2 \right.
+ \left( u^f_n[t]\right)^2 - 2 u^f_{n-1}[t] Q^f_n[t] \\
&& + \left. \left. 2 u^f_n[t] \left( Q^f_n[t] + \pi_{in(f,n)}[t]
- \pi_{out(f,n)}[t] \right)
\right| Q[t] \right].
\end{eqnarray*}

Since $\lambda$ is strictly inside the region $\Lambda,$ there
exist a positive constant $\epsilon$ and a vector of link rates
$\mu$ such that
\[
\mu_{nm} ~\geq~ (1 + \epsilon)\sum_{f: (n,m)\in L(f)} \lambda_f ~,
\quad \mbox{and} \quad \mu ~\in~ co(\Gamma).
\]
Hence,
\begin{eqnarray*}
\sum_{f \in \cF} Q^f_{b(f)}[t] \lambda_f &=&
 \sum_{f \in \cF} \sum_{(n,m) \in L(f)}
\lambda_f \left( Q^f_n[t] - Q^f_m[t] \right) \\
&\leq& \frac{1}{1+\epsilon} \sum_{(n,m) \in \cL} \mu_{nm}
\max_{f:(n,m) \in L(f)} \left( Q^f_n[t] - Q^f_m[t] \right)^+.
\end{eqnarray*}
Therefore,
\begin{eqnarray*}
\dsp \Delta V[t] &\leq& B_1[t]
- \sum_{(n,m)\in\cL} \left( \pi^{*}_{nm}[t] - \mu_{nm}\right)
\max_{f:(n,m) \in L(f)} \left( Q^f_n[t] - Q^f_m[t] \right)^+ \\
&& - ~ \frac{\epsilon}{1 + \epsilon}
\sum_{(n,m) \in \cL} \mu_{nm} \max_{f:(n,m) \in L(f)} \left(
Q^f_n[t] - Q^f_m[t] \right)^+.
\end{eqnarray*}
Now, for any flow $f \in \cF,$ we have that
\begin{eqnarray*}
\sum_{n \in R(f)} Q^f_n[t] &\leq& |R(f)| \sum_{(n,m)\in
L(f)} \left( Q^f_n[t] - Q^f_m[t]\right)^+ \\
&\leq& |R(f)| \sum_{(n,m)\in L(f)} \max_{g: (n,m)\in L(g)}
\left( Q^g_n[t] - Q^g_m[t]\right)^+ \\
&\leq& \frac{K_{max}}{\mu_{L(f)}} \sum_{(n,m) \in \cL}
\mu_{nm} \max_{g:(n,m) \in L(g)} \left( Q^g_n[t] - Q^g_m[t] \right)^+,
\end{eqnarray*}
where $\mu_{L(f)}>0$ is the minimum link rate $\mu_{nm}$ of any
link which is part of the flow's route; obviously,
$\mu_{L(f)}\ge \lambda_f$. Thus, for any flow $f \in \cF,$
\begin{eqnarray}
\Delta V[t] &\leq&
B_1[t] - \frac{\epsilon}{1+\epsilon} \frac{\lambda_f}{K_{max}}
\sum_{n \in R(f)} Q^f_n[t].\label{eqn:drift}
\end{eqnarray}
Note that $B_1[t] \leq b |\cF| K_{max},\, \forall t,$ for some
constant $b>0$ which depends only on $c_{max}$ (see model
definition).
%which is independent of $|\cF|$ and $K_{max}.$
Thus, by manipulating (\ref{eqn:drift}), $\forall f \in \cF,$
we obtain
\[
\limsup_{T\rightarrow\infty}\frac{1}{T}\sum_{t=1}^T\bE \left[ \sum_{n \in R(f)} Q^f_n[t] \right]
~\leq~ \frac{1+\epsilon}{\epsilon} \frac{b}{\lambda_f} |\cF| K_{max}^2.
\]
The above bound along with the positive recurrence of $Q[t]$
gives the desired result.

\section*{Acknowledgments}
The work of the first two authors has been supported in part by DTRA Grant HDTRA1-08-1-0016, NSF
Grant CNS 07-21286, and Army MURI 2008-01733.

\bibliographystyle{IEEEtran}
\bibliography{refs}

\end{document}